\documentclass{article}
\usepackage[a4paper, left=1.2in,right=1.2in,top=1.2in,bottom=1.2in]{geometry}
\usepackage{tikz}
\usepackage[sort, numbers]{natbib}
\usepackage{graphicx}
\usepackage{float}
\usepackage{booktabs}
\usepackage{bbm}
\usepackage{bm}
\usepackage{nicefrac}
\usepackage{ifthen}
\usepackage{algorithm}
\usepackage{algpseudocode}
\usepackage{amsfonts,amsthm}
\usepackage[hyphens]{url}
\usepackage{xspace}
\usepackage{mathtools}
\usepackage{color-edits}
\usepackage[hidelinks]{hyperref}

\addauthor[Sixie]{sy}{red} 
\addauthor[Eugene]{yv}{blue}

\newtheorem{defn}{Definition}[section]

\newtheorem{theorem}{Theorem}

\newtheorem{corollary}[theorem]{Corollary}

\newtheorem{lemma}{Lemma}[section]

\allowdisplaybreaks
\DeclareMathOperator{\argmax}{arg\,max}

\DeclareMathSymbol{\R}{\mathord}{AMSb}{"52}
\DeclarePairedDelimiter\norm{\lVert}{\rVert}

\providecommand{\SET}[1]{\ensuremath{\{ #1 \}}\xspace}
\providecommand{\Set}[2]{\ensuremath{\SET{#1 \mid #2}}\xspace}
\providecommand{\SetCard}[1]{\ensuremath{| #1 |}\xspace}
 
\providecommand{\Trace}[1]{\ensuremath{\text{Tr}\left( #1 \right)}\xspace} 
\providecommand{\Sign}[1]{\ensuremath{\text{sign}\left( #1 \right)}\xspace}

\providecommand{\NumAgent}{\ensuremath{n}\xspace} 
\providecommand{\NumEdge}{\ensuremath{m}\xspace} 
\providecommand{\VSet}{\ensuremath{\mathcal{V}}\xspace} 
\providecommand{\ESet}{\ensuremath{\mathcal{E}}\xspace} 
\providecommand{\MVSet}{\ensuremath{\mathcal{V}^+}\xspace} 
\providecommand{\BVSet}{\ensuremath{\mathcal{V}^-}\xspace} 
\providecommand{\RemoveS}{\ensuremath{\mathcal{S}}\xspace} 
\providecommand{\AdjElem}[2]{\ensuremath{A_{#1, #2}}\xspace} 
\providecommand{\Adj}{\ensuremath{\bm{A}}\xspace} 
\providecommand{\Config}{\ensuremath{\bm{\pi}}\xspace} 
\providecommand{\ConfigBar}{\ensuremath{\bar{\bm{\pi} }}\xspace} 
\providecommand{\ConfigElem}[1]{\ensuremath{\pi_{#1}}\xspace} 
\providecommand{\ConfigElemBar}[1]{\ensuremath{\bar{\pi}_{#1}}\xspace} 
\providecommand{\Loss}{\ensuremath{L}\xspace} 
\providecommand{\TradeOff}[1]{\ensuremath{\alpha_{#1}}\xspace} 
\providecommand{\TradeOffVec}{\ensuremath{\bm{\alpha}}\xspace} 
\providecommand{\DecisionVec}{\ensuremath{\bm{x}}\xspace} 
\providecommand{\Decision}[1]{\ensuremath{x_{#1}}\xspace} 
\providecommand{\DecisionBar}[1]{\ensuremath{\bar{x}_{#1}}\xspace} 
\providecommand{\DecisionVecZ}{\ensuremath{\bm{z}}\xspace} 
\providecommand{\DecisionZ}[2]{\ensuremath{z_{#1, #2}}\xspace} 
\providecommand{\ConfigProb}{\ensuremath{\mathbb{P}}\xspace} 
\providecommand{\ConfigProbEst}{\ensuremath{\hat{\mathbb{P}}}\xspace} 
\providecommand{\MINTLP}{\ensuremath{\texttt{MINT-LP}}\xspace} 
\providecommand{\Simplex}[1]{\ensuremath{\Delta_{#1}}\xspace} 
\providecommand{\Perturb}{\ensuremath{\bm{\delta}}\xspace} 
\providecommand{\PBudget}{\ensuremath{\epsilon}\xspace} 
\providecommand{\MeanEst}{\ensuremath{ \hat{\bm{\mu}} }\xspace} 
\providecommand{\CovEst}{\ensuremath{ \hat{\bm{\Sigma}} }\xspace} 

\providecommand{\Data}{\ensuremath{\mathcal{D}}\xspace} 
\providecommand{\Pred}{\ensuremath{f}\xspace} 
\providecommand{\Feat}{\ensuremath{\bm{x} }\xspace} 

\newcommand{\QMat}{\ensuremath{\bm{Q}}\xspace}  
\newcommand{\bVec}{\ensuremath{\bm{b}}\xspace}  
\newcommand{\QMatHat}{\ensuremath{\hat{\bm{Q}}}\xspace}  
\newcommand{\DecisionVecHat}{\ensuremath{\hat{\bm{x}}}\xspace}  
\newcommand{\XHat}{\ensuremath{\hat{\bm{X}}}\xspace}  
\newcommand{\JMat}{\ensuremath{\bm{J}}\xspace}

\title{Removing Malicious Nodes from Networks\thanks{A preliminary version of this work appeared
   in the Proceedings of the 18th International Conference on
   Autonomous Agents and Multiagent Systems (AAMAS-19)~\cite{yu2019removing}.}}
\author{Sixie Yu\thanks{sixie.yu@wustl.edu}, Yevgeniy Vorobeychik}
\date{ }

\numberwithin{equation}{section}
\begin{document}


\maketitle

\begin{abstract}
A fundamental challenge in networked systems is detection and removal of suspected
malicious nodes.
In reality, detection is always imperfect, and the decision about
which potentially malicious 
nodes to remove must trade off false positives (erroneously removing
benign nodes) and false negatives (mistakenly failing to remove
malicious nodes).
However, in network settings this conventional tradeoff must also
account for node connectivity.
In particular, malicious
nodes may exert malicious influence, so that mistakenly leaving some of
these in the network may cause damage to spread.
On the other hand, removing benign nodes causes direct harm to these,
and indirect harm to their benign neighbors who would wish
to communicate with them.
We formalize the problem of removing potentially malicious nodes from
a network under uncertainty  through an objective that takes connectivity into account.
We show that optimally solving the resulting problem is NP-hard.
We propose a theoretically guaranteed approximation algorithm for solving it based on Semidefinite Programming  relaxation and randomized rounding.
We then further scale our technique to deal with large-scale networks by reformulating the problem as a linear programming.
A combination of both theoretical and empirical analysis, the latter using both
synthetic and real data, provide strong evidence that our algorithmic
approaches are highly effective.
\end{abstract}

\section{Introduction}
    One of the major problems in networked settings is to identify and remove potentially malicious nodes.
For example, malicious nodes in social networks may correspond to accounts created by malicious parties which spread social spam, hate speech, fake news, and the like, with considerable deliterious effects~\cite{allcott2017social,cheng2015antisocial}.
Major social network platforms consequently devote considerable efforts to identifying and removing fake or malicious accounts~\cite{facebook18,nyt2017facebook}.
Nevertheless, evidence suggests that the problem remains pervasive~\cite{vinicius2018brazil,narayanan2018polarization}.
Similarly, in cyber-physical systems (e.g., smart grid infrastructure), computing nodes compromised by malware can cause catastrophic losses, and mitigation through detection and removal of such malicious nodes is a major problem~\cite{mo2012cyber,Yang17}.

A common thread in these scenarios is the tradeoff faced in deciding which nodes to remove: removing a benign node (false positive) causes damage to this node, which may be inconvenience or loss of productivity, and potentially also results in indirect losses to its neighbors; on the other hand, failing to remove a malicious node (false negative) can have deleterious effects as malicious influence spreads to its neighbors. The key observation is that the loss associated with a decision whether to remove a node depends both on the node's likelihood of being malicious and its \emph{local network structure}. Therefore the central challenge faced in deciding which potentially malicious nodes to remove is to account for the combination of uncertainty about whether particular nodes are malicious, and the indirect (network) effects of the decision. This combination makes the decision about which nodes to remove fundamentally a  challenging combinatorial optimization problem. 

We consider the problem of choosing which subset of nodes to remove from a network given an associated probability distribution over joint realizations of all nodes as either malicious or benign (that is, we allow the probability that node $i$ is malicious to depend on whether its neighbors are malicious, as in collective classification and relational learning~\cite{Macskassy07,AMN:taskar2004learning}).
We then model the problem as minimizing the expected loss with respect to this distribution, where the loss function is composed of three parts: the \emph{direct} loss ($\Loss_1$) stemming from removed benign nodes, the \emph{indirect} loss associated with cutting links between removed and remaining benign nodes ($\Loss_2$), and the loss associated with malicious nodes that remain, quantified in terms of \emph{links} these have to benign nodes ($\Loss_3$). 

\begin{figure}[ht]
\centering
\setlength{\tabcolsep}{0.1pt}
\begin{tabular}{c}
\includegraphics[width=0.6\columnwidth]{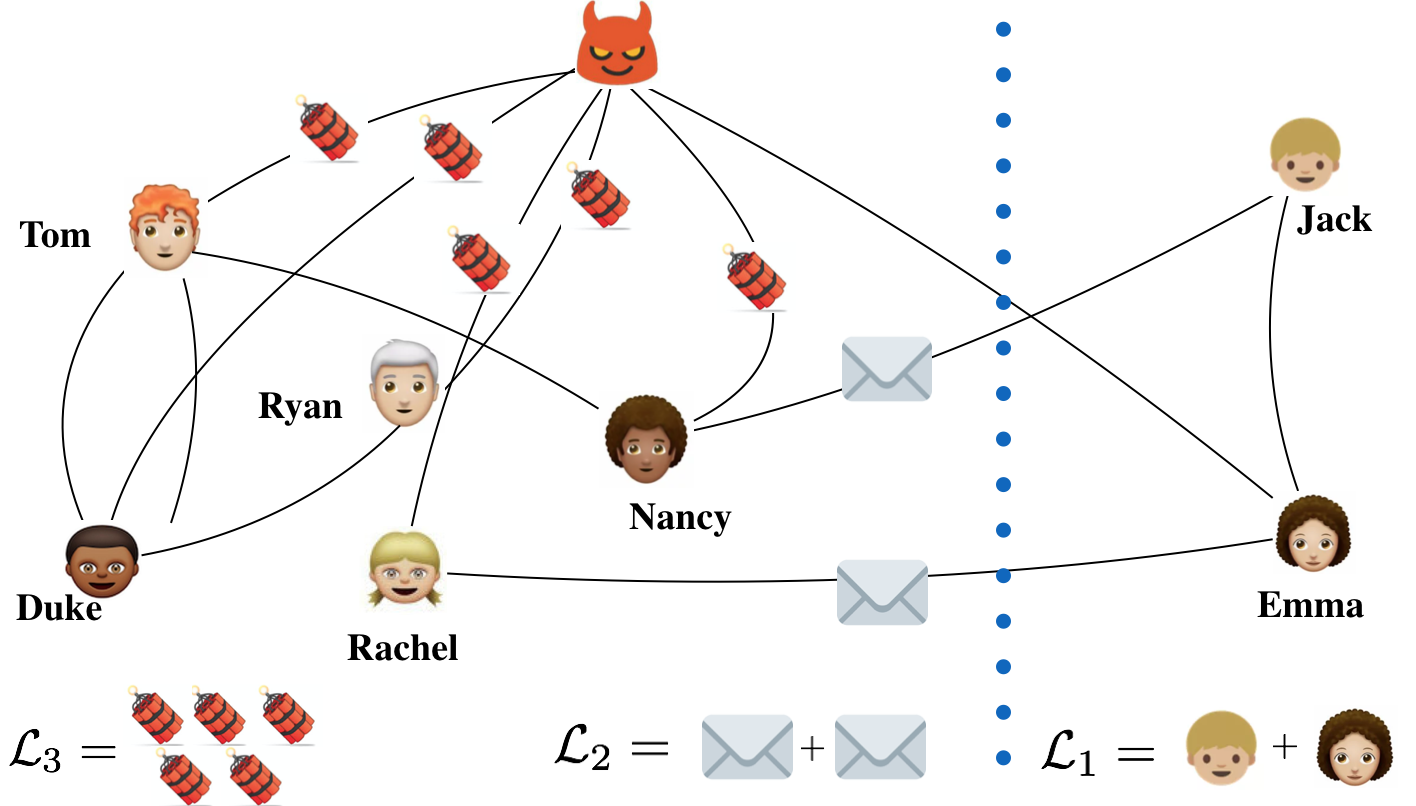} 
\end{tabular}
\caption{An illustration of a decision to remove two nodes, Jack and Emma, from the network, on our loss function.}
\label{fig:loss_demo}
\end{figure}

To illustrate, consider Figure~\ref{fig:loss_demo}.
In this example, we have decided to remove Jack and Emma, the two \emph{benign} nodes to the right of the vertical dotted line.
On the other hand, we chose not to remove the malicious node in red.
Suppose that we pay a penalty of $\TradeOff{1}$ for each benign node we remove, a penalty of $\TradeOff{2}$ for each link we cut between two benign nodes, and $\TradeOff{3}$ for each link between \emph{remaining} malicious and benign nodes.
Since we removed $2$ benign nodes ($\Loss_1 = 2$), cut $2$ links between benign nodes (one between Jack and Nancy, and another between Emma and Rachel; $\Loss_2 = 2$), and the malicious node is still connected to $5$ benign nodes (Tom, Duke, Ryna, Rachel, and Nancy; $\Loss_3 = 5$), our total loss is $2\TradeOff{1} + 2\TradeOff{2} + 5\TradeOff{3}$.
If we instead only removed the malicious node, our total loss would have been $0$, while removing the malicious node instead of Emma (but together with Jack) would result in the loss of $\TradeOff{1} + 2\TradeOff{2}$.

Our first contribution is an efficient algorithmic approach for choosing a subset of nodes to remove by approximately minimizing the expected loss function described above, where the expectation is taken with respect to an exogenously specified node maliciousness distribution for a given network; the distribution can be estimated by any suitable machine learning model, e.g., logistic regression, Markov network, or graph neural network.
However, we show that  minimizing the loss is NP-hard, even with highly restrictive settings. 
Our solution strategy is resorting to convex relaxation, i.e., we relax the original combinatorial optimization problem to a convex Semidefinite programming (SDP) in continuous domain, which can be solved in polynomial time.
We solve the convex relaxation for a globally optimal solution; then we convert it to an approximate solution in the discrete domain by utilizing a principled randomized rounding procedure.  
Finally, we show a theoretical guarantee on the quality of the approximate solution.
For simplicity, we call the approximation algorithm the SDP-based algorithm.

The scalability of the SDP-based algorithm is limited by its computational costs.
To improve the scalability, we reformulate the original optimization problem as a linear programming (LP) with standard linearization and relaxation techniques; the resulting LP can easily handle instances with thousands of nodes.
Due to the improvement on scalability, we are able to experiment on an extensive array of instances parameterized by combinations of the trade-off parameters, i.e., $\TradeOff{1}, \TradeOff{2}$ and $\TradeOff{3}$.
One surprising observation is that the LP formulations of most instances admit integral solutions (hence optimal) despite the NP-hardness; in other words, most instances are ``easy'', in the sense that they are solvable in polynomial time; in contrast, we call those instances without integral solutions ``hard. 
Based on the observation, we design an LP-based heuristic algorithm to solve the problem; the idea is to slightly perturb the trade-off parameters of a ``hard'' instance such that the perturbed instance can be optimally solved by the LP.

\noindent{\bf Related Work } 
There are several lines of research that are related to the problem of choosing which subset of nodes to remove from a network. 
The first relevant literature involves the problem of \emph{network node classification} or \emph{collective classification}~\cite{AMN:taskar2004learning,Macskassy07,sen2008collective,KipfW17}.
The fundamental task in this line of research is to assign labels to  nodes in a network.
In this approach, one makes use of the attributes associated with the nodes as well as the connections among them to learn a joint distribution \ConfigProbEst over the assignment of labels to nodes.
Viewing this problem from the perspective of identifying malicious nodes, a typical next step would be to make the most likely assignment of labels to nodes, which involves the use of inference algorithms (e.g., the sum-product algorithm for computing a maximum a posterior estimate), itself a non-trivial problem \cite{wainwright2008graphical}.
Collective classification methods would be an input into our approach, in the form of the learned probability distribution \ConfigProbEst; however, as our experiments demonstrate, our approach is significantly better than the naive baseline which removes nodes that are most likely to be malicious, as it is crucial to account for the network structure not merely at inference time, but also at decision time.

Another related line of research considers a problem of \emph{graph scan statistics} and hypothesis testing~\cite{arias2011detection,priebe2005scan,sharpnack2013near}. These study the following problem: given a graph $G$ where each node is associated with a random variable with an exogenously specified probability distribution, find a subset of nodes that maximizes a scan statistic defined over subsets of nodes (for example, this statistic may generalize log-likelihood ratio). \citet{arias2011detection} proposed a scan statistic for special graph models. \citet{priebe2005scan} proposed a scan statistic defined over clusters with special geometric structures. These methods do not easily generalize to arbitrary graph models or arbitrary scan statistics. \citet{sharpnack2013near} employed the generalized log-likelihood ratio as the scan statistic. By assuming that the set of malicious nodes has sparse connections with others, the hypothesis test can be converted to solving a graph cut problem, which is further relaxed into a convex optimization by leveraging the Lov{\'a}sz extension of a graph cut. This method is not scalable to large graphs and it does not account for potentially inaccurate estimates of the node maliciousness distribution.

Our problem is loosely related to the burgeoning field of adversarial machine learning~\cite{Vorobeychik18}, although we do not explicitly consider issues of adversarial response in our setting, as well as the broader problem of security games on networks~\cite{alpcan2010network}. 
There are several approaches that explicitly consider adversarial response in a networked environment. \citet{yu2018adversarial} consider how to mitigate the spread of malicious information on social networks by modeling the problem as a Stackelberg game between the defender, who aims to mitigate the spread of malicious information, and an attacker who tries to initiate malicious spread. \citet{hajaj2019adversarial} conducted behavioral experiments to investigate the impact of adversarial behavior in a networked coordination task. \citet{vorobeychik2015securing} formulated the problem of securing a set of interdependent assets  as a Stackelberg security game, where interdependence can be abstracted as a graph $G$. Finally, a number of approaches study the specific problem of subverting social network analysis techniques such as link prediction \cite{waniek2018attack,zhou2019attacking} and node classification~\cite{dai2018adversarial,zugner2018adversarial}.

The remaining of the paper is organized as follows:
    \begin{enumerate}
        \item In Section~\ref{ch2:model}, we introduce \ref{eq:MINT}, a novel model and formulation of optimal removal of malicious nodes on networks under uncertainty by reducing from the \textsc{Maximum Independent Set} problem; the reduction implies that there is no constant-factor approximation algorithm exists.
        We show that optimally solving \ref{eq:MINT} is NP-hard.
        \item In Section~\ref{ch2:algo}, we first describe the SDP-based approximation algorithm and the associated theoretical guarantee on the quality of approximation.
        We then describe the LP-based heuristic algorithm and empirically verify its efficacy on large-scale networks.
        \item In Section~\ref{ch2:ret}, we provide experimental results to show that both algorithms outperform the alternatives on an large array of instances.
    \end{enumerate}

\section{Model}\label{ch2:model}

Consider a network $G=(\VSet, \ESet)$, where $\VSet$ ($|\VSet|=\NumAgent$) is the set of nodes and $\ESet$ the set of edges. 
Each node $i \in \VSet$ represents a user and each edge $(i, j)$ represents some connection (e.g., friendship) between user $i$ and user $j$. 
We assume that $G$ is undirected and  has no  self-loop.
The adjacency matrix \Adj of the graph is symmetric, with zeros on the diagonal.
The entries of $\Adj$ are binary if the graph is unweighted (i.e., $\AdjElem{i}{j} \in \SET{0, 1}$), or some non-negative real numbers if the graph is weighted (i.e., $\AdjElem{i}{j} \in \R_+$). 
Subsequent presentation focuses on unweighted graphs; generalization to weighted graphs is straightforward.

We explain the problem by first considering complete information about the identity of malicious and benign nodes, and subsequently describe our actual model in which this information is unknown. 
Specifically, let $\Config \in \SET{0, 1}^\NumAgent$ be a \emph{configuration} of the network, with $\ConfigElem{i}=1$ indicating that a node $i$ is malicious, with $\ConfigElem{i}=0$ when $i$ is benign. 
For notational convenience, we also define the complement $\ConfigElemBar{i} = 1 - \ConfigElem{i}$, i.e., $\ConfigElemBar{i}=1$ if and only if $i$ is benign.  
Consequently, $\Config$ (and $\ConfigBar$) assigns malicious or benign labels to each node. 
Let the set of malicious and benign nodes be $\MVSet$ and $\BVSet$, respectively. 
Our goal is to remove a subset $\RemoveS$ of nodes such that the impact of the remaining malicious nodes on the network is minimized, while limiting disruptions caused to the benign subnetwork.

To formalize this intuition, we define a loss function associated with the set $\RemoveS$. 
This loss function has three components, each corresponding to a key consideration in the problem.
The first part of the loss function is defined below:
    \begin{equation}
        \Loss_1 = \left | \BVSet \cap \RemoveS \right |, 
    \end{equation}
which is the cardinality of wrongly removed nodes, encoding the \emph{direct} loss associated with removing benign nodes; this simply penalizes every false positive, as one would naturally expect.
The loss $\Loss_1$ ignores the broken \emph{relationships} among benign nodes that result from the decision, which is captured by the second component $\Loss_2$ as follows:
    \begin{equation}
        \Loss_2=\Big| 
            \Set{ (i, j) \in \ESet }{ 
                i \in  \left( \BVSet \cap (\VSet \setminus \RemoveS) \right)  , 
                j \in  \left( \BVSet \cap \RemoveS \right)}
        \Big|.
    \end{equation}
The loss $\Loss_2$ imposes a penalty for cutting connections between benign nodes that are removed (i.e., $\BVSet \cap \RemoveS$) and benign nodes that remain (i.e., $\BVSet \cap (\VSet \setminus \RemoveS)$).
In other words, $\Loss_2$ captures the \emph{indirect} consequence of removing benign nodes \emph{on the structure of the benign subnetwork}.
This aspect is critical to capture in network settings, as relationships and connectivity are what networks are about.
The third component $\Loss_3$ measures the consequence of \emph{failing to remove malicious nodes} in terms of connections from these to benign nodes:
    \begin{equation}
        \Loss_3=\Big | 
            \Set{ (i,j) \in \ESet}{ 
                i \in  \left( \MVSet \cap (\VSet \setminus \RemoveS) \right) , 
                j \in  \left( \BVSet \cap (\VSet \setminus \RemoveS) \right)}
        \Big|.        
    \end{equation}
The sets $\MVSet \cap (\VSet \setminus \RemoveS)$ and $\BVSet \cap (\VSet \setminus \RemoveS)$ represent the malicious and the benign nodes that remain in the network, respectively.
Intuitively, we can think of $\Loss_3$ as encoding remaining deleterious effect that the malicious nodes still exert on the benign nodes, e.g., spammers keep sending annoying emails to normal users.

The total loss combines these three components as a weighted sum, i.e., $\Loss = \TradeOff{1} \Loss_1 + \TradeOff{2} \Loss_2 + \TradeOff{3} \Loss_3$, with $\TradeOff{1} + \TradeOff{2} + \TradeOff{3} = 1$ and each $\TradeOff{i}$ being nonnegative.
Other than this constraint, we allow $\TradeOff{i}$s to be arbitrary relative weights of the different components, specified depending on the domain.
For example, if we are concerned about false positives (i.e., wrongly removing benign nodes), but not very much about network structure, we would set $\TradeOff{1} \gg \TradeOff{2}$.
Alternatively, we would set $\TradeOff{3}$ to be very large (compared with $\TradeOff{1}$ and $\TradeOff{2}$) to indicate that the loss of false negatives dominates the others. 

We now rewrite the loss function in a way that will prove more mathematically convenient. 
Define a vector $\DecisionVec \in \SET{0, 1}^\NumAgent$, where $\Decision{i}=1$ if and only if node $i$ is removed ($i \in \RemoveS$), and $\Decision{i}=0$ if node $i$ remains in the network ($i \in \VSet \setminus \RemoveS$). 
Recall that the goal of the model is to identify a subset of nodes $\RemoveS$ to remove so as to minimize the impact of the remaining malicious nodes on the network, while  limiting disruptions caused to the benign subnetwork. 
This goal is naturally captured by the loss function below:

    \begin{equation}\label{eq:loss}
        \begin{aligned}
            \Loss(\DecisionVec) = 
                \TradeOff{1}    \underbrace{\sum_{i=1}^{\NumAgent}{  \Decision{i} \ConfigElemBar{i}  }}_{  \Loss_1  }   + 
                \TradeOff{2} 
                \underbrace{\sum_{i,j}^{\NumAgent}{  \AdjElem{i}{j}  \Decision{i} \DecisionBar{j} \ConfigElemBar{i} \ConfigElemBar{j}   }}_{  \Loss_2  }   + 
                \TradeOff{3}    \underbrace{\sum_{i,j}^{\NumAgent}{  \AdjElem{i}{j} \DecisionBar{i} \DecisionBar{j}   \ConfigElem{i} \ConfigElemBar{j}   } }_{\Loss_3}.
        \end{aligned}
    \end{equation}

With complete information, it is immediate that the loss is minimized if $\RemoveS$ contains all, and only, the malicious nodes. 
In reality, the identity of malicious and benign nodes is unknown to us, and instead we have a probability distribution  over these. 
This probability distribution may be obtained by learning probability that a node is malicious given its features from past data, e.g., with a Markov network~\citep{AMN:taskar2004learning} or a graph neural network~\citep{KipfW17}.
To formalize, let $\Config \sim \ConfigProbEst$, where $\ConfigProbEst$ captures the joint probability distribution over node configurations (malicious or benign).
For our purposes, we make no assumptions on the nature of this distribution; a special case would be when maliciousness probabilities for nodes are independent (conditional on a node's observed features), but our model also captures natural settings in which configurations of network neighbors are correlated (e.g., when malicious nodes tend to have many benign neighbors).  
The loss function described in \eqref{eq:loss} becomes the following expected loss:
        \begin{equation}\label{eq:expected_loss}
            \begin{aligned}
                  \mathbb{E}_{\Config \sim \ConfigProbEst}[\Loss(\DecisionVec)] = 
                  \TradeOff{1} \sum_{i=1}^{\NumAgent}{     \Decision{i}   \mathbb{E}[\ConfigElemBar{i}]     }   + 
                  \TradeOff{2} \sum_{i,j}^{\NumAgent}{     \AdjElem{i}{j} \Decision{i} \DecisionBar{j} \mathbb{E}[\ConfigElemBar{i} \ConfigElemBar{j}]     } + 
                  \TradeOff{3} \sum_{i,j}^{\NumAgent}{     \AdjElem{i}{j} \DecisionBar{i} \DecisionBar{j} \mathbb{E}[\ConfigElem{i} \ConfigElemBar{j}]     }.
            \end{aligned}
        \end{equation}

To have a compact formulation we re-write the expected loss in matrix notations.
We first define auxiliary matrices $\bm{B}, \bm{P}$ and $\bm{M}$ as follows 
\footnote{$\text{diag}(\bm{x})$ returns a diagonal matrix with $\bm{x}$ as the diagonal entries.}:
        \begin{equation}\label{eq:auxiliary-matrix}
                \begin{aligned}
                    \bm{B} & = \text{diag}\big( \mathbb{E}[\ConfigBar] \big)\\
                    \bm{P} & = \Adj \odot \mathbb{E}[\ConfigBar \ConfigBar^\top] \\
                    \bm{M} & = \Adj \odot \mathbb{E}[ \Config \ConfigBar^\top],
                \end{aligned}
        \end{equation}
where $\odot$ is the element-wise multiplication.
Let $\MeanEst \in \R^\NumAgent$ and $\CovEst \in \R^{\NumAgent \times \NumAgent}$ be the mean and covariance of \ConfigProbEst.
With some algebraic transformations, the matrices  $\bm{P}$ and $\bm{M}$ are functions of the mean and covariance, i.e.,
    \begin{equation}
        \begin{aligned}
             \bm{P} & = \Adj \odot \left( \JMat - \bm{1} \MeanEst^\top - \MeanEst \bm{1}^\top + \CovEst + \MeanEst \MeanEst^\top \right) \\
             \bm{M} & = \Adj \odot \left(\MeanEst \bm{1}^\top - \CovEst - \MeanEst\MeanEst^\top \right),
        \end{aligned}
    \end{equation}
where $\bm{1} \in \R^\NumAgent$ (resp. $\JMat \in \R^{\NumAgent \times \NumAgent}$) is the all-ones vector (resp. matrix).
The expected loss $\mathbb{E}_{\Config \sim \ConfigProbEst}[\Loss(\DecisionVec)]$ is re-written with the matrix notations as follows
    \begin{equation}\label{eq:qp}
             \DecisionVec^\top (\TradeOff{3}\bm{M} -    \TradeOff{2}\bm{P})\DecisionVec 
            + 
              \DecisionVec^\top (\TradeOff{1}\bm{B}\bm{1} + \TradeOff{2}\bm{P}\bm{1} - \TradeOff{3}\bm{M}\bm{1} - \TradeOff{3}\bm{M}^\top \bm{1})
            + 
              \TradeOff{3}\bm{1}^\top \bm{M} \bm{1}.
    \end{equation}
Our model, which we term \ref{eq:MINT}, is now captured by the following optimization problem with constants omitted: 

        \begin{equation}\label{eq:MINT}\tag{\texttt{MINT}}
            \begin{aligned}
                & \min_{ \DecisionVec } & & \DecisionVec^\top \QMat \DecisionVec + \DecisionVec^\top \bVec  \\
                &s.t.     &    &    \DecisionVec \in \SET{0, 1}^\NumAgent,
            \end{aligned}
        \end{equation}
where $\QMat = \left( (\TradeOff{3}\bm{M} - \TradeOff{2}\bm{P}) + (\TradeOff{3}\bm{M} - \TradeOff{2}\bm{P})^\top \right) / 2$ is a real symmetric matrix and $\bVec =\TradeOff{1}\bm{B}\bm{1} + \TradeOff{2}\bm{P}\bm{1} - \TradeOff{3}\bm{M}\bm{1} - \TradeOff{3}\bm{M}^\top \bm{1}$.
        
\ref{eq:MINT} is a \emph{Quadratic Programming} (QP).
Optimally solving a QP is usually difficult, even when \QMat has a single negative eigenvalue~\cite{pardalos1991quadratic}; in fact, it has been shown that even finding a local optimum  is NP-hard~\cite{ahmadi2020complexity}.
However, there are a few cases where optimally solving a QP is tractable.
For example, when \QMat is positive semidefinite and \DecisionVec is in continuous space, the resulting QP is a convex Semidefinite programming (SDP) and can be solved in polynomial time.
Another example does not require the positive semidefiniteness of \QMat; instead, 
the decision variable needs to be continuous and there is a single constraint quadratic in \DecisionVec; in this case strong duality holds if the Slater's condition is satisfied and we can solve the dual of the original QP (see Appendix B at~\cite{boyd2004convex}).
An extensive line of research is devoted to study QP and SDP.
In particular, approximation algorithms based on SDP have attracted many research interests (\cite{lovasz1979shannon,goemans1995improved,frieze1997improved,nesterov2000semidefinite} and the references therein).

Notice that \ref{eq:MINT} is motivated from a binary classification perspective, i.e., each node is identified either benign or malicious.
This perspective is natural in many applications, e.g., anomaly detection.
However, \ref{eq:MINT} is \emph{not} limited to binary settings, rather, it can be extended to multi-class scenarios (i.e., $\Decision{i} \in \SET{0, 1, \ldots, k-1}$) for $k$-class problems by leveraging the ideas described in~\cite{kleinberg2002approximation}.

\subsection{Computational Hardness}\label{ch2:hardness}

We now show that optimally solving \ref{eq:MINT} is NP-hard, even for highly restrictive settings.
The reduction is from the \textsc{Maximum Independent Set} (MIS) problem.

        \begin{theorem}\label{ch2:hard}
            Minimizing \ref{eq:MINT} is NP-hard even when
                \begin{itemize}
                    \item the probability that a node is malicious restricts to $1/2$, i.e., $\mathbb{E}_{\Config \sim \ConfigProbEst}[\ConfigElem{i}]=1/2$ for all $i$, and
                    \item the random variables \ConfigElem{i} and \ConfigElem{j} are independent for any $i \ne j$.
                \end{itemize}
        \end{theorem}
\begin{proof}
\newcommand{\KSet}{\ensuremath{\mathcal{K}}\xspace}  
\newcommand{\KSetP}{\ensuremath{\mathcal{K}}\xspace} 
\newcommand{\BSet}{\ensuremath{\mathcal{K}}\xspace}  

Given a graph $G=(\VSet, \ESet)$, the MIS problem is to find an independent set in $G$ of maximum cardinality.
We consider the special case of \ref{eq:MINT} with the restrictions listed in Theorem~\ref{ch2:hard}; in addition, we ignore the loss of cutting edges among benign nodes (i.e., $\TradeOff{2}=0$), and let the loss of wrongly keeping malicious nodes in the benign subnetwork dominate the others (i.e., $\TradeOff{3} \gg \TradeOff{1}$).
The loss function of the special case is as follows:
    \begin{equation}
        \Loss^\prime = \underbrace{\frac{\TradeOff{1}}{2} \sum_{i=1}^{\NumAgent}{\Decision{i}}}_{\Loss_1^{\prime}} + \underbrace{\frac{\TradeOff{3}}{4}\sum_{i,j}^{\NumAgent}{\AdjElem{i}{j} \DecisionBar{i} \DecisionBar{j} }}_{\Loss_3^{\prime}}.
    \end{equation}
Let \KSet be the maximum independent set of $G$, and \RemoveS the set of removed nodes.
We first show that keeping only \KSet is the optimal solution, i.e, removing the set $\RemoveS =  \VSet \setminus \KSet$ is optimal.
Notice that removing any node from \KSet is sub-optimal, as an extra $\frac{\TradeOff{2}}{2}$ loss is incurred to $\Loss_1^{\prime}$.  
Next, we show that it is sub-optimal to put any node in \RemoveS back to \KSet.
Suppose we put a set of nodes $\BSet \subseteq \RemoveS$ back to \KSet; this must introduce additional edges to \KSet, otherwise it is not the maximum independent set. 
As $\TradeOff{3} \gg \TradeOff{1}$, introducing any edge to \KSet is sub-optimal.
Since we cannot add or remove any node from  \KSet,  we should keep the nodes in \KSet and remove the set $\RemoveS = \VSet \setminus \KSet$.

For the other direction,  suppose  $\RemoveS = \VSet \setminus \KSet$ minimizes $\Loss^\prime$.
We show that \KSet is the maximum independent set of $G$.
First, suppose \KSet is not an independent set, which means that there is at least a pair of nodes connected by an edge. 
As $\TradeOff{3} \gg \TradeOff{1}$, removing the nodes decreases $\Loss^\prime$,  which contradicts that $\RemoveS$ minimizes $\Loss^\prime$.
Thus, \KSet must be an independent set. 
Next, we show that \KSet is maximum. 
Suppose there exists another set independent set \KSetP and $\SetCard{\KSetP} > \SetCard{\KSet}$. 
Then keeping the nodes in \KSetP would further decrease $\Loss^\prime$ by decreasing $\Loss_1^\prime$, which contradicts the fact that \RemoveS minimizes the $\Loss^\prime$. 
We conclude that \KSet is the maximum independent set.
\end{proof}

It is known that no constant-factor approximation algorithm exists for solving the MIS problem unless P=NP (Section 4.4 at \cite{trevisan2004inapproximability}).
As a result, we have that no constant-factor approximation algorithm exists for solving \ref{eq:MINT}.
    \begin{corollary}\label{cor:no-const-mint}
        No constant-factor approximation algorithm exists for solving \ref{eq:MINT} unless P=NP.
    \end{corollary}

\section{Algorithm}\label{ch2:algo}
    
\subsection{SDP-based Approximation Algorithm}\label{ch2:algo-sdp}
In this section we present an approximation algorithm to solve \ref{eq:MINT}.
The general idea of the algorithm is to relax the binary quadratic programming (BQP) to a convex SDP.
Then we solve the SDP for a global optimum, which can be done in polynomial time by using, e.g., the interior point method~\cite{boyd2004convex}.
The global optimum is in continuous space; we convert it to a binary solution of \ref{eq:MINT} by utilizing a classic randomized rounding procedure~\cite{goemans1995improved}.
In what follows, we first describe the convex relaxation and subsequently present the randomized rounding procedure.     

\subsubsection{Convex Relaxation}\label{ch2:convex-relax}
We first convert the feasible region to $\SET{-1, 1}^\NumAgent$ by substituting the original decision variable \DecisionVec
with $(\DecisionVec + \bm{1}) / 2$.
The resulting objective function is
    \begin{equation}
        \left(      \frac{\DecisionVec+ \bm{1}}{2}      \right)^\top \QMat 
        \left(      \frac{\DecisionVec+ \bm{1}}{2}      \right) 
        + 
        \left(      \frac{\DecisionVec+ \bm{1}}{2}      \right)^\top \bVec.
    \end{equation}
Omitting constant terms,  the optimization problem of \ref{eq:MINT} becomes
        \begin{equation}\label{eq:MINT_nonhomo}
            \begin{aligned}
                &\min_{\DecisionVec}         &    & \frac{1}{4} \DecisionVec^\top \QMat \DecisionVec + \frac{1}{2} \DecisionVec^\top (\QMat \bm{1} + \bVec) \\
                &s.t.     &    &    \DecisionVec \in \{ -1,1\}^\NumAgent.
            \end{aligned}
        \end{equation}
In subsequent discussion, we assume that $\DecisionVec^\top(\QMat\bm{1} + \bVec) \ne 0$; this is to 
exclude the pathological case where both $\DecisionVec$ and the complement $-\DecisionVec$ achieve the same loss.
The problem in \eqref{eq:MINT_nonhomo} is a \emph{non-homogeneous} quadratic programming due to the linear term involving \DecisionVec (i.e., $\frac{1}{2}\DecisionVec^\top(\QMat\bm{1} + \bVec)$).
For convenience of later analysis, we reformulate \eqref{eq:MINT_nonhomo} to a  \emph{homogeneous} quadratic programming without the linear term.
The idea is to augment the original formulation with an extra dimension.
First, define a real symmetric matrix $\QMatHat \in \mathbb{S}^{\NumAgent + 1}$ 
    \begin{equation}
        \QMatHat = 
        \begin{bmatrix}
            (1/4)\QMat &  (1/4)\left(\QMat \bm{1} + \bVec \right) \\
            (1/4)\left( \bm{1}^\top \QMat + \bVec^\top \right) & 0
        \end{bmatrix}.
    \end{equation}
Then, let $\DecisionVecHat \in \SET{-1, 1}^{\NumAgent + 1}$ be the augmented decision variable, i.e., $\DecisionVecHat=\begin{bmatrix} \DecisionVec^\top, t \end{bmatrix}^\top$ with $t \in \SET{-1, 1}$.
With the introduction of \QMatHat and \DecisionVecHat, the homogeneous formulation is as follows 
    \begin{equation}\label{eq:MINT_opt}
        \begin{aligned}
            & \min_{\DecisionVecHat} & & \DecisionVecHat^\top \QMatHat \DecisionVecHat \\
            &s.t.     &    &    \DecisionVecHat \in \SET{-1, 1}^{\NumAgent+1}.
        \end{aligned}
    \end{equation}
    
\noindent{\bf Remark }
It is well-known that the decision version of \eqref{eq:MINT_opt} is NP-complete~\cite{garey1979computers}; in fact, this is still true even if we assume that \QMatHat is positive semidefinite, since we can add a large constant of identity to the objective function without changing the optimal solution; in other words, the problem does not become easier even if the objective function is convex; see the discussion in Section 2.2 of \cite{blekherman2012semidefinite}.

Next, we convert the homogeneous formulation to an SDP, a convex programming solvable in polynomial time.
The key step is to introduce an auxiliary matrix $\XHat =\DecisionVecHat \DecisionVecHat^\top$ and utilize the following relations (see Section 2.2 at \cite{blekherman2012semidefinite}):
        \begin{equation}\label{ch2:trace_rela}
            \DecisionVecHat^\top \QMatHat \DecisionVecHat = \Trace{\DecisionVecHat^\top \QMatHat \DecisionVecHat} = \Trace{\QMatHat \DecisionVecHat \DecisionVecHat^\top} = \Trace{\QMatHat \XHat},
        \end{equation}
where $\Trace{\cdot}$ is the trace operator; the second equality is because the trace operator is invariant under cyclic permutations.
With \eqref{ch2:trace_rela},  the optimization problem in \eqref{eq:MINT_opt} is reformulated as follows:
        \begin{equation}\label{ch2:nonconvex-tmp}
            \begin{aligned}
            & \min_{ \XHat, \DecisionVecHat } & & \Trace{\QMatHat \XHat} \\
            &s.t.     &    &    \XHat = \DecisionVecHat  \DecisionVecHat^\top \\
            &         &    & \DecisionVecHat \in \SET{-1, 1}^{\NumAgent+1}.
            \end{aligned}
        \end{equation}
Notice that the diagonal entries of \XHat are all ones.
In addition, by the definition of \XHat, it follows that $\bm{z}^\top \DecisionVecHat \DecisionVecHat^\top \bm{z} = \left( \DecisionVecHat^\top \bm{z} \right)^\top \left( \DecisionVecHat^\top \bm{z} \right) \ge 0$ for any $\bm{z} \in  \R^{\NumAgent + 1} $.
Thus, the matrix \XHat is positive semidefinite. 
A classic approach to relax the above optimization to an SDP is by replacing the equality constraint with the following~\cite{luo2010semidefinite}:
        \begin{equation}
            \XHat \succeq \bm{0} \text{ and }  \hat{X}_{ii} = 1, i=1, \ldots, \NumAgent+1,
        \end{equation}
where $\XHat \succeq \bm{0}$ means that \XHat is positive semidefinite.
It is direct to verify that any feasible solution of \eqref{ch2:nonconvex-tmp} satisfies the above constraints.
Finally, the convex relaxation of \eqref{eq:MINT_opt} is as follows:
        \begin{equation}\label{eq:MINT_SDP}
            \begin{aligned}
                & \min_{\XHat \in \mathbb{S}^{\NumAgent+1}} & & \Trace{\QMatHat \XHat}\\
                &s.t.     &    &    \hat{X}_{ii} = 1, i=1,\ldots, \NumAgent+1 \\
                &         &    &    \XHat \succeq 0.
            \end{aligned}
        \end{equation}
        
We solve \eqref{eq:MINT_SDP} for an optimal solution $\XHat^\ast$, which can be done in polynomial time by using the interior point method~\cite{boyd2004convex}. 
The solution $\XHat^\ast$ is a real symmetric matrix in continuous space; however, what we ultimately need is a discrete solution in $\SET{-1, 1}^\NumAgent$.
Next, we describe a randomized rounding procedure to generate such a discrete solution from $\XHat^\ast$.

\subsubsection{Randomized Rounding}\label{S:random}

\newcommand{\VMat}{\ensuremath{\bm{V}}\xspace}  
\newcommand{\VVec}{\ensuremath{\bm{v}}\xspace}  
\newcommand{\LMat}{\ensuremath{\bm{L}}\xspace}  
\newcommand{\DMat}{\ensuremath{\bm{D}}\xspace}  
\newcommand{\ZVec}{\ensuremath{\bm{z}}\xspace}  

The convex relaxation described in the previous section leads to a lower bound on the optimal value of \ref{eq:MINT}, however, it offers no guidance about constructing a feasible solution of \ref{eq:MINT} with small loss.
In this section, we describe a randomized procedure to fill this gap; the procedure is based on the famous Goemans-Williamson rounding~\cite{goemans1995improved}.

Let  $\XHat^\ast$ be the optimal solution of the SDP (i.e., \eqref{eq:MINT_SDP}). Due to the positive semi-definiteness, the matrix $\XHat^\ast$ admits a \emph{Cholesky Decomposition} 
\footnote{
    The Cholesky Decomposition is usually for \emph{positive definite} matrices. For a positive semidefinite matrix $\XHat^\ast$,  we first apply the \textit{LDL Decomposition}, which results in $\XHat^\ast=\LMat \DMat \LMat^\top$.
    The matrix $\DMat$ is a diagonal matrix with nonnegative diagonal entries.
    The Cholesky decomposition of $\XHat^\ast$ is then $\VMat^\top \VMat$ with $\VMat = \DMat^{\frac{1}{2}}\LMat^\top$.
}.
Specifically, the decomposition has the form $\XHat^\ast=\VMat^\top \VMat$, where $\VMat \in \R^{r\times (\NumAgent+1)}$ and $r$ is the rank of $\XHat^\ast$.  
Let the $i$-th column of $\VMat$ be $\VVec_i$.
The $(i, j)$-th element of $\XHat^\ast$ is $\VVec^\top_i \VVec_j$. 
The constraints in \eqref{eq:MINT_SDP} enforce that $\hat{X}^\ast_{ii} = 1$ for all $i$.
Consequently, we have $\hat{X}^\ast_{ii} = \VVec_i^\top \VVec_i = \norm{\VVec_i}^2_2 =1$, i.e., $\VVec_i$ are on the surface of the unit sphere centered at the origin.
We now utilize the vectors $\VVec_1, \ldots, \VVec_{\NumAgent+1}$ to generate a feasible solution of \eqref{eq:MINT_opt}.
Let $\ZVec \in \R^r$ be a vector sampled uniformly at random from the unit sphere\footnote{This can be achieved by normalizing a standard multivariate Gaussian random variable, i.e., sample $\bm{z} \sim \mathcal{N}(\bm{0}, \bm{1})$ and then normalize $\bm{z} / \norm{\bm{z}}_2^2$.}.
The feasible solution is obtained with the following rules:
        \begin{equation}
            \Decision{i} = \begin{cases}
		            \Sign{ \mathbb{E}_{\ZVec}[\ZVec^\top \VVec_i] }, &  \text{if $t = +1$} \\
		            -\Sign{ \mathbb{E}_{\ZVec}[\ZVec^\top \VVec_i] }, & \text{if $t = -1$},
	              \end{cases}
        \end{equation}
where $t=\Sign{ \mathbb{E}_{\ZVec}[\ZVec^\top \VVec_{\NumAgent+1}]  }$; it is only used to determine the signs of $\Decision{i}$.
The randomness of the expectations comes from the sampling of $\bm{z}$.
We summarize the randomized rounding procedure in Algorithm~\ref{algo:random}. 
After obtaining $\DecisionVec \in \SET{-1, 1}^\NumAgent$, the nodes with $\Decision{i}=1$ 
are removed from the network while the nodes with $\Decision{i}=-1$ remain.

In summary, the approximation algorithm to solve \ref{eq:MINT} consists of 1) solving the SDP in \eqref{eq:MINT_SDP}  and 2) converting the solution of the SDP to a feasible solution $\DecisionVec \in \SET{-1, 1}^\NumAgent$ by a randomized rounding procedure; the approximation algorithm is described in Algorithm~\ref{algo:MINT}.

\begin{algorithm}[ht]
\caption{Randomized Rounding}\label{algo:random}
\begin{algorithmic}[1]
\State   \textbf{Input}: The optimal solution $\XHat^\ast \in \mathbb{S}^{\NumAgent+1}$ of the SDP (i.e., \eqref{eq:MINT_SDP})
        \State Cholesky Decomposition to obtain $\XHat^\ast=\VMat^\top \VMat$, where the columns of $\VMat$ are denoted by $\VVec_1,\ldots, \VVec_{\NumAgent+1}$ 
        \State Sample $\bm{z} \sim \mathcal{N}(\bm{0}, \bm{1})$ and then normalize, i.e., $\bm{z} / \norm{\bm{z}}_2^2$
        \State Compute $t=\Sign{ \mathbb{E}_{\ZVec}[\ZVec^\top \VVec_{\NumAgent+1}]  }$
        \State Generate a feasible solution $\DecisionVec \in \SET{-1, 1}^\NumAgent$  of \eqref{eq:MINT_opt} as follows:
        \begin{equation*}
            \Decision{i} = \begin{cases}
		            \Sign{ \mathbb{E}_{\ZVec}[\ZVec^\top \VVec_i] }, &  \text{if $t = +1$} \\
		            -\Sign{ \mathbb{E}_{\ZVec}[\ZVec^\top \VVec_i] }, & \text{if $t = -1$}.
	              \end{cases}
        \end{equation*}
		\State \textbf{Return}: $\DecisionVec = \SET{\Decision{1}, \ldots, \Decision{\NumAgent}}$
\end{algorithmic}
\end{algorithm}

\begin{algorithm}[htbp]
\caption{SDP-based Approximation Algorithm to Solve \ref{eq:MINT}} \label{algo:MINT}
\begin{algorithmic}[1]
\State \textbf{Input}: $\QMat$, $\bVec$
		\State Construct  $\QMatHat$ as follows:  
        \begin{equation*}
            \QMatHat = 
            \begin{bmatrix}
                (1/4)\QMat &  (1/4)\left(\QMat \bm{1} + \bVec \right) \\
                (1/4)\left( \bm{1}^\top \QMat + \bVec^\top \right) & 0
            \end{bmatrix}.
        \end{equation*}
        \State Solve the SDP (i.e., \eqref{eq:MINT_SDP}) for the optimal solution $\XHat^\ast$ 
        \State Apply the randomized rounding Algorithm~\ref{algo:random} to get $\DecisionVec = \SET{\Decision{1}, \ldots, \Decision{\NumAgent}}$
        \State $\RemoveS=\Set{i}{\Decision{i}=1, i=1,\ldots, \NumAgent}$ is the set of nodes to remove
\end{algorithmic}
\end{algorithm}

\subsubsection{Performance Guarantee}
\newcommand{\DMatHat}{\ensuremath{\hat{\bm{D}}}\xspace}  
\newcommand{\PMatHat}{\ensuremath{\hat{\bm{P}}}\xspace}  

In this section we show a performance guarantee for the approximation algorithm. 
Recall that the algorithm consists of two steps. 
The first step is to solve the SDP; the second step is to convert the optimal solution of the SDP to a feasible solution $ \DecisionVec \in \SET{-1, 1}^\NumAgent$. 
The derivation of the guarantee is inspired by the fact that the $\DecisionVec$  enjoys nice approximation guarantees if the matrix \QMatHat admits some special structures.
One particular structure is called \emph{diagonally dominant matrix}, which is defined below:
        \begin{defn}\label{def:diag-mat}
            A real  symmetric matrix $\bm{P} \in \mathbb{S}^{n}$ is called diagonally dominant if $P_{ii} \ge \sum_{j\ne i}^{}{|P_{ij}|}$ for all $i=1,\ldots, n$.
        \end{defn}

We briefly introduce the usefulness of diagonally dominant matrices when bounding the approximation quality of \DecisionVec.\footnote{See Chapter 2 of~\cite{blekherman2012semidefinite} for detailed discussion.}
To be consistent with the literature we switch to maximization problems; it is direct to transfer the results back to a minimization problem with a sign change of the objective function.

Consider the optimization problems listed in Table~\ref{tab:bqp-sdp}, where the matrix $\bm{P}$ is a symmetric matrix.
The left cell is a typical Binary Quadratic Programming (BQP), which includes as special cases our problem \ref{eq:MINT}, as well as other classic combinatorial optimization problems, e.g., \textsc{MAXCUT} problem.
The middle cell is the corresponding SDP of the BQP.
The optimal objective value of the SDP is an upper bound on that of the BQP, i.e., $v_{SDP}^\ast \ge v^\ast$.
Given the optimal solution of the SDP, let \DecisionVec be a feasible solution generated from the Goemans-Williamson rounding; its objective value is denoted with $v(\DecisionVec)=\DecisionVec^\top \bm{P} \DecisionVec$.
It is direct that $v^\ast \ge v(\DecisionVec)$ since an arbitrary solution in $\SET{-1, 1}^\NumAgent$ is a lower bound on $v^\ast$.
However, we would like $v(\DecisionVec)$ to be close to $v^\ast$ in some sense.

The right cell of Table~\ref{tab:bqp-sdp} shows two cases where $v(\DecisionVec)$ is close to $v^\ast$ in expectation.
When the matrix $\bm{P}$ is diagonally dominant (d.d.), \citet{goemans1995improved} showed that the ratio of $\mathbbm{E}[v(\DecisionVec)]$ over $v^\ast$ is at least $0.878$.
A similar ratio ($\approx 0.636$) holds when $\bm{P}$ is positive semidefinite (p.s.)~\cite{nesterov1998semidefinite}.

\begin{table}[ht]
\centering
\begin{tabular}{c|c|c}
\hline
BQP                                                                                                                                                                                                                            & The SDP Relexation                                                                                                                                                                                                                                                                                       & Approx. Guarantees                                                                                                                                                                                                                          \\
\begin{tabular}[c]{@{}c@{}}$\begin{aligned}\\                 & \max_{\DecisionVec} & & \DecisionVec^\top \bm{P} \DecisionVec \\\\                 &s.t.     &    &    \DecisionVec \in \SET{-1, 1}^\NumAgent,\\             \end{aligned}$\end{tabular} & \begin{tabular}[c]{@{}c@{}}$\begin{aligned}\\             & \max_{\XHat \in \mathbb{S}^\NumAgent} & &  \Trace{\bm{P} \XHat} \\\\             &s.t.     &    &    \XHat \succeq 0 \\\\             &         &    &    \hat{X}_{ii} = 1, i=1,\ldots, \NumAgent.\\             \end{aligned}$\end{tabular} & \begin{tabular}[c]{@{}c@{}}$\begin{aligned}\\ & \text{d.d.}: 0.878 v^\ast_{SDP} \le \mathbbm{E}[v(\DecisionVec)] \\\\ & \text{p.s.}:0.636 v^\ast_{SDP} \le \mathbbm{E}[v(\DecisionVec)] \\ \end{aligned}$\end{tabular} \\ \hline
\end{tabular}
\caption{\textbf{Left}: a typical binary quadratic programming; \textbf{Middle}: the corresponding SDP relaxation; \textbf{Right}: the approximation guarantees when $\bm{P}$ is diagonally dominant (d.d.) or positive semidefinite (p.s.).}
\label{tab:bqp-sdp}
\end{table}

The approximation guarantee when $\bm{P}$ is a symmetric diagonally dominant matrix is formally stated in Theorem~\ref{lemma:SDP_bound}.

    \begin{lemma}[\cite{blekherman2012semidefinite}, Section 2.2.2]\label{lemma:SDP_bound}
        Consider the following binary quadratic programming:
        \begin{equation}\label{eq:P}\tag{\texttt{P}} 
            \begin{aligned}
                & \max_{\DecisionVec} & & \DecisionVec^\top \bm{P} \DecisionVec \\
                &s.t.     &    &    \DecisionVec \in \SET{-1, 1}^\NumAgent,
            \end{aligned}
        \end{equation}
        where $\bm{P}$ is a symmetric diagonally dominant matrix. 
        The SDP relaxation of problem \eqref{eq:P} is as follows:
        \begin{equation}\label{eq:P_SDP}\tag{\texttt{SDP}} 
            \begin{aligned}
            & \max_{\XHat \in \mathbb{S}^\NumAgent} & &  \Trace{\bm{P} \XHat} \\
            &s.t.     &    &    \XHat \succeq 0 \\
            &         &    &    \hat{X}_{ii} = 1, i=1,\ldots, \NumAgent.
            \end{aligned}
        \end{equation}
        Let the objective values of \eqref{eq:P} and \eqref{eq:P_SDP} at optimality be $v^{\ast}$ and $v^{\ast}_{SDP}$, respectively. 
        In addition, let $\DecisionVec \in \SET{-1, 1}^\NumAgent$ be the feasible solution of \eqref{eq:P} generated from the Goemans-Williamson rounding.
        The following relation holds with $\alpha \approx 0.878$:
        \begin{equation}\label{eq:SDP_bound}
            0 \le \alpha \cdot v^{\ast}_{SDP} \le \mathbbm{E}[\DecisionVec^\top \bm{P} \DecisionVec] \le v^{\ast} \le v^{\ast}_{SDP}.
        \end{equation}
    \end{lemma}

The nonnegativity of $v^{\ast}$ (and hence $v^{\ast}_{SDP}$) follows from the fact that a real symmetric diagonally dominant matrix with nonnegative diagonal entries is positive semidefinite~\cite{ddm}.
Based on Lemma~\ref{lemma:SDP_bound},  we give an approximation guarantee to Algorithm~\ref{algo:MINT}.
The key idea is to convert \eqref{eq:MINT_opt} to a BQP involving a diagonally dominant matrix.
It is worth nothing that the guarantee is not a constant-factor bound due to the hardness result in Corollary~\ref{cor:no-const-mint}.

        \begin{theorem}\label{th:guarantee}
            Let $V^\ast$ and $V^\ast_{SDP}$ be the optimal objective values of \eqref{eq:MINT_opt}  and its SDP, respectively.
            Let $\DecisionVecHat \in \SET{-1, 1}^{\NumAgent+1}$ be the solution generated from  Algorithm~\ref{algo:random}.
            The following relations hold:
            \begin{equation}\label{eq:guarantee}
                V^{\ast}_{SDP} \le V^{\ast} \le \mathbbm{E}[\DecisionVecHat^\top \QMatHat \DecisionVecHat]  \le \alpha\cdot V^{\ast}_{SDP}  + (1-\alpha) \cdot c,
            \end{equation}
            where $\alpha \approx 0.878$ and $c = \sum_{i}^{\NumAgent+1}{ \sum_{j\ne i}^{\NumAgent+1}{|\hat{Q}_{ij}|}  }$.
        \end{theorem} 
        \begin{proof}

        To be consistent with the notations used in Lemma~\ref{lemma:SDP_bound}, we convert the minimization problem of \eqref{eq:MINT_opt}  to the  following maximization problem:

        \begin{equation}\label{eq:MINT_max}
            \begin{aligned}
                & \max_{\DecisionVecHat} & & -\DecisionVecHat^\top \QMatHat \DecisionVecHat \\
                &s.t.     &    &    \DecisionVecHat \in \SET{-1, 1}^{\NumAgent + 1}.
            \end{aligned}
        \end{equation}

        Note that matrix \QMatHat is usually \emph{not} a diagonally dominant matrix;  the diagonal entries are all zeros, however, some off-diagonal entries may be nonzero.
        Thus, we cannot directly apply Lemma~\ref{lemma:SDP_bound}. 
        To resolve this issue, 
        we add an instance-dependent constant  $c = \sum_{i}^{\NumAgent+1}{ \bigg( \sum_{j\ne i}^{\NumAgent+1}{|\hat{Q}_{ij}|}  \bigg) }$ to the objective function, such that it becomes a quadratic function involving a symmetric diagonally dominant matrix.
        Specifically, let $\DMatHat$ be a diagonal matrix with $\hat{D}_{ii}=\sum_{j\ne i}^{}{|\hat{Q}_{ij}|}$. 
        Since $\hat{x}_i^2=1$, we express the constant $c$ as a quadratic term involving \DecisionVecHat, i.e., $c = \DecisionVecHat^\top \DMatHat \DecisionVecHat$.
        Thus,  the objective function with the constant $c$ becomes:
        \begin{equation}
            \begin{aligned}
                - \DecisionVecHat^\top  \QMatHat \DecisionVecHat + \DecisionVecHat^\top \DMatHat \DecisionVecHat  =  \DecisionVecHat^\top \big( \DMatHat - \QMatHat \big) \DecisionVecHat.
            \end{aligned}
        \end{equation}
        The addition of $c$ does not change the optimal solutions;
        however, it affects the quality of the approximation algorithm, which depends on specifics of the input instance as encoded by \QMatHat.
        Define $\PMatHat = \DMatHat - \QMatHat$ as follows:
        \begin{equation}
            \PMatHat = 
            \begin{bmatrix} 
                    \sum_{j\ne 1}^{\NumAgent+1}{|\hat{Q}_{1j}|} & -\hat{Q}_{1,2} &  \dots  & -\hat{Q}_{1, \NumAgent+1} \\ 
                     -\hat{Q}_{21} & \sum_{j\ne 2}^{\NumAgent+1}{|\hat{Q}_{2j}|} & \dots &   -\hat{Q}_{2, \NumAgent+1} \\ 
                    \vdots & \vdots & \dots  & \vdots  \\
                    -\hat{Q}_{\NumAgent+1, 1} & -\hat{Q}_{\NumAgent+1, 2} & \dots & \sum_{j\ne \NumAgent+1}^{\NumAgent+1}{|\hat{Q}_{\NumAgent+1, j}|}  
           \end{bmatrix}.
        \end{equation}
        It is direct to verify that \PMatHat is a symmetric diagonally dominant matrix.
        Consider the following optimization problem \ref{eq:MINT_plus}, with the ``+'' indicating that the problem is a maximization problem:
        \begin{equation}\label{eq:MINT_plus}\tag{\texttt{MINT+}}
            \begin{aligned}
                & \max_{\DecisionVecHat} & & \DecisionVecHat^\top \PMatHat \DecisionVecHat \\
                &s.t.     &    &    \DecisionVecHat \in \SET{-1, 1}^{\NumAgent+1}.
            \end{aligned}
        \end{equation}

        As mentioned before, the problem \ref{eq:MINT_plus} has the same optimal solutions as Eq.~\eqref{eq:MINT_max}.
        The SDP relaxation of \ref{eq:MINT_plus} is readily obtained.
        Let $v_+^{\ast}$ and $v_{SDP+}^{\ast}$ be the optimal objective values of \ref{eq:MINT_plus} and the SDP, respectively.
        Also, let $\DecisionVecHat$ be the feasible solution of \eqref{eq:MINT_plus} generated from Algorithm~\ref{algo:random}.
        A direct application of  Lemma~\ref{lemma:SDP_bound} leads to the following with $\alpha \approx 0.878$.
        \begin{equation}\label{eq:const_bound}
            \alpha \cdot v_{SDP+}^{\ast} \le \mathbbm{E}[\DecisionVecHat^\top \PMatHat \DecisionVecHat] \le v_+^{\ast} \le v_{SDP+}^{\ast}.
        \end{equation}

        Let the optimal objective values of \eqref{eq:MINT_opt} and its SDP (i.e., Eq.~\eqref{eq:MINT_SDP}) be $V^\ast$ and $V^\ast_{SDP}$, respectively.
        Observe that $v^{\ast}_+ = c - V^{\ast}$, $v^{\ast}_{SDP+} = c - V^{\ast}_{SDP}$ and $\mathbbm{E}[\DecisionVecHat^\top \PMatHat \DecisionVecHat] = c - \mathbbm{E}[\DecisionVecHat^\top \QMatHat \DecisionVecHat]$; substituting these back to Eq.~\eqref{eq:const_bound} leads to
            \begin{equation}
                V^{\ast}_{SDP} \le V^{\ast} \le \mathbbm{E}[\DecisionVecHat^\top \QMatHat \DecisionVecHat]  \le \alpha\cdot V^{\ast}_{SDP}  + (1-\alpha) \cdot c.
            \end{equation}
        \end{proof}

Notice that the constant $c$ in Theorem~\ref{th:guarantee} is not universal.
Instead, it is \emph{instance-dependant} as it is a function of \QMatHat.
Thus, it is natural to expect that some instances admit small values of $c$ while others do not; this affects the tightness of the upper bound on $\mathbbm{E}[\DecisionVecHat^\top \QMatHat \DecisionVecHat]$.

\subsection{LP-based Approximation Algorithm}\label{ch2:algo-lp}

Although the SDP-based approximation algorithm enjoys nice theoretical guarantees, it is challenging to handle large-scale networks due to computational costs.
In particular, Algorithm~\ref{algo:MINT} involves solving an SDP, which is  computationally expensive; the general method to solve an SDP is the \emph{primal-dual interior point method}, which under mild conditions converges to a pre-defined accuracy in $O(\NumAgent^{1/2})$ Newton iterations; each iteration involves roughly $O(\NumAgent^3)$ operations (see Section 5.7 of~\cite{SDP-opt:vandenberghe1996semidefinite}).

In this section, we describe another algorithm that is scalable to large-scale networks.
Specifically, we relax \ref{eq:MINT} to a linear programming (LP) by standard linearization and relaxation techniques.
One surprising observation is that for most instances of \ref{eq:MINT}  the resulting LP admits integral solutions;  this indicates that  most instances of \ref{eq:MINT} are easy in the sense that they are solvable in polynomiam time.
We leverage this observation and propose an LP-based algorithm to solve \ref{eq:MINT}.

\subsubsection{Linearization and Relaxation of \ref{eq:MINT}}
The high-level idea is to transfer the expected loss $\mathbb{E}[\Loss(\DecisionVec)]$ (i.e., Eq.~\eqref{eq:expected_loss}) from a binary quadratic function to a linear function involving binary variables.
Then we relax the binary variables to continuous domain (i.e., $[0, 1]^\NumAgent$).
A detailed introduction of the linearization and relaxation is in \cite{LP-opt:glover1973further}.

First, we linearize each quadratic term in the expected loss by introducing a new variable $\DecisionZ{i}{j}= \Decision{i} \Decision{j}$.
Specifically, we add three constraints to guarantee that $\DecisionZ{i}{j} = 1$ \emph{if and only if} both $\Decision{i}$ and $\Decision{j}$ are equal to $1$, i.e., 
    \begin{equation}
        \begin{aligned}
        &\DecisionZ{i}{j} \le \Decision{i} \\
        &\DecisionZ{i}{j} \le \Decision{j} \\
        &\DecisionZ{i}{j} \ge \Decision{i} + \Decision{j} - 1.
        \end{aligned}
    \end{equation}
The relaxation step is simply replacing the original feasible region with the continuous region $[0, 1]^\NumAgent$.
The resulting LP is as follows. 

\begin{equation}\label{eq:MINT-LP}\tag{\texttt{MINT-LP}}
\begin{aligned}
& \min_{\DecisionVec \in [0,1]^\NumAgent, \DecisionVecZ \in [0,1]^{\SetCard{\ESet}} } & & \sum_{i,j}^{\NumAgent}{\Gamma_{ij}\left( \Decision{i} - \DecisionZ{i}{j} \right)} + \sum_{j=1}^{\NumAgent}{\phi_j \Decision{j} }\\
&s.t.     &    &    0 \le \Decision{i} \le 1, \,\, \text{for all $i$} \\
&         &    &    0 \le \DecisionZ{i}{j} \le 1, \,\, \text{for all $(i, j) \in \ESet$} \\
&         &    &    \DecisionZ{i}{j} \le \Decision{i}, \,\, \text{for all $(i, j) \in \ESet$} \\
&         &    &    \DecisionZ{i}{j} \le \Decision{j}, \,\, \text{for all $(i, j) \in \ESet$} \\
&         &    &    \DecisionZ{i}{j} \ge \Decision{i} + \Decision{j} - 1, \,\, \text{for all $(i, j) \in \ESet$}.
\end{aligned}
\end{equation}
Recall that $\ESet$ is the edge set of $G$.
The coefficient $\Gamma_{ij}$ is the $(i, j)$-th entry of the matrix $\bm{\Gamma}=\TradeOff{2} \bm{P} - \TradeOff{3} \bm{M}$, and $\phi_j$ is the $j$-th element of the vector $\bm{\phi}= \TradeOff{1} \bm{B} \bm{1} - \TradeOff{3}\bm{M}^\top \bm{1}$, where $\bm{P}, \bm{M}$ and $\bm{B}$ are defined in \eqref{eq:auxiliary-matrix}.
A quadratic term $\Decision{i} \Decision{j}$ is in the expected loss if and only if there is an edge between $i$ and $j$; thus, the dimension of \DecisionVecZ is $\NumEdge$---the number of edges in $G$.
The total number of variables is $n+m$, and the total number of constraints is $2n + 2m + 3m$.
Therefore, the size of \ref{eq:MINT-LP} is linear with respect to the size of $G$.

\subsubsection{Integral Solutions of \ref{eq:MINT-LP}}
An interesting observation is that many instances of \ref{eq:MINT-LP} admit integral solutions, i.e., the optimal solutions consist of precisely $1$ or $0$.
Intuitively, this means that the hardness result of \ref{eq:MINT} may be too pessimistic and the worst-case instances can be rare in practice.
To verify the intuition, we identify an instance of \MINTLP with \TradeOffVec and  \ConfigProbEst, i.e., $\MINTLP(\TradeOffVec, \ConfigProbEst)$; the vector \TradeOffVec includes the trade-off parameters, i.e., $\TradeOffVec=(\TradeOff{1}, \TradeOff{2}, \TradeOff{3})$; the distribution \ConfigProbEst is the estimated joint distribution over the maliciousness of nodes.
We assume a fixed \ConfigProbEst and characterize the ``hard'' and ``easy'' instances in the space of \TradeOffVec, which gives us an informal picture of where the ``hard'' instances reside.

\begin{figure}[ht]
\def\FigSize{2in}
\centering
\setlength{\tabcolsep}{0.01pt}
\begin{tabular}{ccc}
\includegraphics[width=\FigSize]{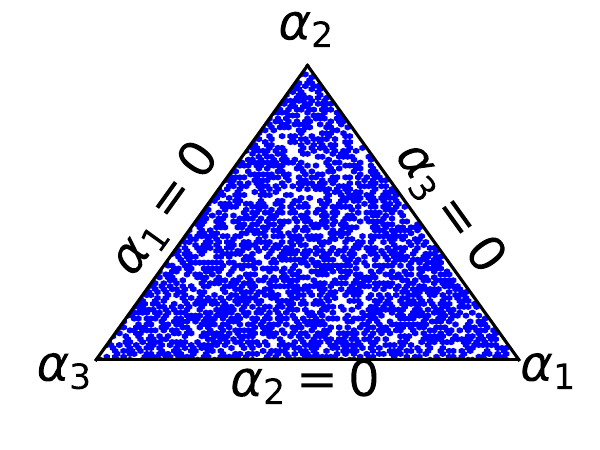} &
\includegraphics[width=\FigSize]{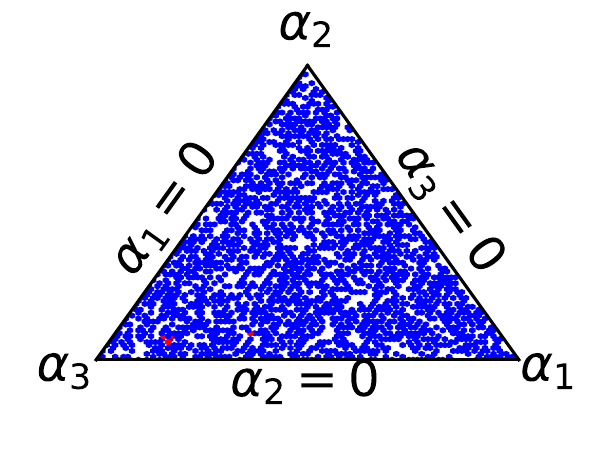} & 
\includegraphics[width=\FigSize]{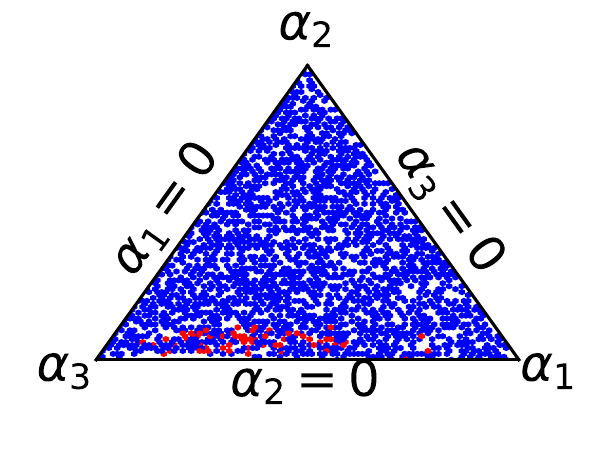}
\end{tabular}
\caption{Instances $\MINTLP(\TradeOffVec, \ConfigProbEst)$ generated by sampling \TradeOffVec from \Simplex{2}. 
Each subplot corresponds to a particular underlying network with a specific \ConfigProbEst. 
From left to right, Barab\'asi-Albert, Watts-Strogatz, and a subgraph of the Facebook network. 
Blue dots (resp. red dots) are the instances admitting integral (resp. fractional) solutions.}
\label{fig:LP-ins}
\end{figure}

Let $\Simplex{2}=\Set{\TradeOffVec \ge \bm{0}}{\TradeOff{1} + \TradeOff{2} + \TradeOff{3} = 1}$ be the standard probability simplex.
We generate different instances of \MINTLP by sampling the trade-off parameters \TradeOffVec from \Simplex{2}.
Figure~\ref{fig:LP-ins} visualizes the generated instances, with each subplot corresponds to a particular underlying network structures; each dot represents an instance with a specific trade-off vector, e.g., the top vertices represent the instanecs with $(\TradeOff{1}=0, \TradeOff{2}=1, \TradeOff{3}=0)$.
The subplots (from left to right) correspond to a synthetic Barab\'asi-Albert network~\cite{barabasi1999emergence}, a synthetic Watts-Strogatz network~\cite{watts1998collective}, and a  subgraph of the real-world Facebook network~\cite{mcauley2012learning}; each network is associated with a distinct joint distribution \ConfigProbEst simulated from data.

One surprising observation from Figure~\ref{fig:LP-ins} is that most instances (i.e., the blue dots) admit integral solutions; in other words, they are solvable in polynomial time despite the NP-hardness.
On the other hand, the red dots represent the instances with fractional solutions; for simplicity we call these instances \emph{fractional instances}.
It is clear that fractional instances constitute a small part of the space of instances.
In particular, for the Barab\'asi-Albert network there is no fractional instance at all, in other words, all instances can be solved in polynomial time.
This supports the intuition: most instances of \ref{eq:MINT}  encountered in practice are ``easy'', in the sense that  they are solvable in polynomial time by resorting to the corresponding LP formulation.

\subsubsection{Perturbation-based Algorithm}
With the observation from Figure~\ref{fig:LP-ins}, we describe a scalable algorithm based on \MINTLP.
Given an input graph $G$ and the joint distribution \ConfigProbEst of maliciousness over nodes, we use $\MINTLP(\TradeOffVec)$ to represent an input instance parameterized by the trade-off parameters.
The first step is to solve the corresponding LP, which results in an optimal solution $(\DecisionVec^\ast, \DecisionVecZ^\ast)$. 
If $\DecisionVec^\ast$ is integral, it is also an optimal solution of \ref{eq:MINT}.
If $\DecisionVec^\ast$ contains fractional entries, we slightly perturb \TradeOffVec with a vector $\Perturb \in \R^3$, i.e., $\TradeOffVec^\prime = \TradeOffVec + \Perturb$;
we control the magnitude of \Perturb with some $\ell_p$ norm and a  parameter \PBudget (i.e., $\norm{\Perturb}_p \le \PBudget$) such that the perturbed trade-off vector is close to the original one.
Intuitively, the small perturbation may ``nudge'' the fractional instance ($\MINTLP(\TradeOffVec)$) to a nearby one ($\MINTLP(\TradeOffVec^\prime)$) that admits integral solutions.
For the instances where the perturbation does not find an easy-to-solve alternative, we obtain an integral solution by  randomly rounding the fractional $\DecisionVec^\ast$, e.g., run a Bernoulli trial for the $i$-th node with success probability $\Decision{i}^\ast$.
The algorithm is described in Algorithm~\ref{algo:MINT-LP}.

\begin{algorithm}[ht]
\caption{Scalable \ref{eq:MINT}}\label{algo:MINT-LP}
\begin{algorithmic}[1]
\State   \textbf{Input}: $\TradeOffVec, \bm{\Gamma}, \bm{\phi}, G, \PBudget, \tau$ \Comment{we perturb the original instance at most $\tau$ times} 
        \State Solve \MINTLP for $(\DecisionVec^\ast, \DecisionVecZ^\ast)$
        \If{$\DecisionVec^\ast$ is integral}
            \State \textbf{Return} $\DecisionVec^\ast$
        \Else
            \State $i = 0$
            \While{$i < \tau$}
                \State Generate $\Perturb$ such that $\norm{\Perturb}_p \le \PBudget$ and $\TradeOffVec^\prime = \TradeOffVec + \Perturb \ge 0$ \label{MINT-LP:op1}
                \State $(\hat{\DecisionVec}^\ast, \hat{\DecisionVecZ}^\ast) \leftarrow  \text{solve } \MINTLP(\TradeOffVec^\prime)$
                \If{$\hat{\DecisionVec}^\ast$ is integral}
                    \State \textbf{Return} $\hat{\DecisionVec}^\ast$
                \Else
                    \State $i = i + 1$
                \EndIf
            \EndWhile
            \State Generate a set $\mathcal{R}$ of integral solutions by randomly rounding $\hat{\DecisionVec}^\ast$ \label{MINT-LP:op2}
            \State  $\DecisionVec^\ast \leftarrow \argmax_{\DecisionVec \in \mathcal{R}} \mathbb{E}[\Loss(\DecisionVec)]$ \Comment{$\mathbb{E}[\Loss(\DecisionVec)]$ is defined in Eq.~\eqref{eq:expected_loss}}
            \State \textbf{Return} $\DecisionVec^\ast$
        \EndIf
\end{algorithmic}
\end{algorithm}


\section{Experimental Results}\label{ch2:ret}
    \subsection{Results for the SDP-based Algorithm}
In this section we present experiments to show the effectiveness of \ref{eq:MINT}.
In particular, we compared the performance of the SDP-based approximation algorithm (i.e., Algorithm~\ref{algo:MINT}) with two baselines. 
We focused on synthetic network structures, but derived distribution over maliciousness of nodes (i.e., \ConfigProbEst) using real data.
Specifically, we considered two types of network structures: Barabasi-Albert (BA)~\cite{barabasi1999emergence} and Watts-Strogatz networks (Small-World)~\cite{watts1998collective}. 
BA is characterized by its power-law degree distribution, where the probability that a randomly selected node has $k$ neighbors is proportional to $k^{-r}$.
For the BA network we experimented with three variants, BA-1, BA-2, and BA-3, which differ in the value of the exponent $r$ of their power-law degree distributions. 
For Small-World networks we also experimented with three variants, SW-1, SW-2, and SW-3, that have different local clustering coefficients. For both networks we generated $20$ instances each with $\NumAgent=128$ nodes.

In our experiments, we consider a simplified case where the maliciousness probabilities for nodes are independent. 
In addition,  we assume that a single estimator (e.g., logistic regression ) was trained  to estimate the probability that a node is malicious based on features from past data. 
Note that these assumptions are reasonable for the purpose of validating the effectiveness of our model, since the focus of our model is not how to estimate maliciousness probabilities. 
For more complex cases, for example, when maliciousness probabilities for nodes are correlated, more advanced techniques (e.g., graph neural networks or Markov networks), can be applied to estimate the maliciousness probabilities, but our general approach would not change.

In all of our experiments, we derived \ConfigProbEst from data as follows.
We start with a dataset $\Data$ which includes malicious and benign instances (the meaning of these designations is domain specific), and split it into three subsets: $\Data_{\text{train}}$ , $\Data_1$, and $\Data_{\text{test}}$, with the ratio of $0.3:0.6:0.1$.
Our first step is to learn a probabilistic predictor $\hat{\Pred}(\Feat)$ of maliciousness  from $\Data_{\text{train}}$.
Next, we randomly assign malicious and benign feature vectors from $\Data_{\text{test}}$ to the nodes on the network, assigning $10\%$ of nodes with malicious and $90\%$ with benign feature vectors.
For node $i$, we use its assigned feature vector $\Feat_i \in \Data_{\text{test}}$ to obtain the \emph{estimated} probability of this node being malicious, i.e., $\hat{\Pred}(\Feat_i)$.
The distribution \ConfigProbEst is thus $\Set{\hat{\Pred}(\Feat_i)}{i \in \VSet, \Feat_i \in \Data_{\text{test}}}$, which is used to construct the input to the algorithms.
However, to ensure that our evaluation is fair and reasonably represents realistic limitations of the knowledge of the true maliciousness distribution, we train another probabilistic predictor, $\Pred(\Feat)$, now using the data $\Data_{\text{train}} \cup \Data_1$.
Applying this new predictor to the nodes and their assigned feature vectors, we now obtain a distribution $\ConfigProb=\Set{\Pred(\Feat_i)}{i \in \VSet, \Feat_i \in \Data_{\text{test}}}$ which we use to evaluate performance.

We conducted two sets of experiments. 
In the first set of experiments we used synthetic networks and used data from the Spam~\cite{lichman2013uci} dataset to simulate the  \ConfigProbEst and \ConfigProb, as described above. 
The Spam dataset \Data consists of spam and non-spam instances along with their corresponding labels. 
In the second set of experiments we simulated \ConfigProbEst and \ConfigProb from the Hate Speech data~\cite{davidson2017automated} collected from Twitter.
The Hate Speech dataset is a crowdsourced dataset that contains three types of tweets: 1) hate speech tweets that express hatred against a targeted group of people; 2) offensive language tweets that appear to be rude, but do not explicitly promote hatred; and 3) normal tweets that neither promote hatret nor are offensive. 
We categorized this dataset into two classes in terms of whether a tweet represents Hate Speech, with the offensive language tweets categorized as non-Hate Speech. 
After categorization, the total number of tweets is $24783$, of which $1430$ are Hate Speech. 
We applied the same feature extraction techniques as \citet{davidson2017automated} to process the data.

\noindent{\bf Baselines }
We compared Algorithm~\ref{algo:MINT} with LESS, a state-of-the-art approach for graph hypothesis testing, as well as a simple baseline which removes a node $i$ if its maliciousness probability $\hat{\Pred}(\Feat_i) > \theta^{\ast}$, where 
$\theta^{\ast}$ is a specified threshold.
The algorithm LESS was proposed by \citet{sharpnack2013near}, and considers a related hypothesis testing problem. 
The null hypothesis is that each node in the graph is associated with a random variable sampled from the standard Gaussian $\mathcal{N}(0,1)$, while the alternative hypothesis is that there is a fraction of nodes where the random variables associated with them are sampled from $\mathcal{N}(\mu, 1)$ with $\mu$ other that $0$ (in our interpretation, these are the malicious nodes). 
The algorithm LESS employs the generalized log-likelihood over a subset of nodes as a test statistic, and the hypothesis test is to find the subset that has the strongest evidence against the null hypothesis. 
We remove the subset of nodes found by LESS. 
The simple baseline has a trade-off parameter $\alpha$ between false-positive rate (FPR) and false-negative rate (FNR)  (in our experiments $\alpha=0.5$).  
We select an optimal threshold $\theta^{\ast}$ that minimizes $\alpha \text{FPR} + (1-\alpha) \text{FNR}$ on training data.

\noindent{\bf Results }
The decision of which subset $\RemoveS \subseteq \VSet$ to remove was obtained by running Algorithm~\ref{algo:MINT} and the baseline approaches.
When evaluating the decision, we used the expected loss (i.e., Eq.~\eqref{eq:expected_loss}) and instantiated the underlying distribution as \ConfigProb.
The loss incurred from decision \RemoveS is $\mathbb{E}_{\ConfigProb}[\DecisionVec_{\RemoveS}]$, where $\DecisionVec_\RemoveS \in \SET{0, 1}^\NumAgent$ is the characteristic vector of \RemoveS.
The averaged losses
for experiments where \ConfigProbEst and \ConfigProb were simulated from Spam (resp. Hate Speech) data are shown in Table~\ref{tab:spam} (resp. Table~\ref{tab:hate}). 
The top table contains the results on BA networks and the bottom table contains the results on Small-World networks.
Each row corresponds to a combination of trade-off parameters $(\TradeOff{1},\TradeOff{2},\TradeOff{3})$; for example, ``1-2-7'' corresponds to $(\TradeOff{1}=0.1, \TradeOff{2}=0.2, \TradeOff{3}=0.7)$ and ``equal'' to $(\TradeOff{1}=1/3, \TradeOff{2}=1/3, \TradeOff{3}=1/3)$.
We experimented with four combinations of these: $(0.1, 0.2, 0.7)$, $(0.2, 0.7, 0.1)$, $(0.7, 0.2, 0.1)$, and $(1/3, 1/3, 1/3)$. 
Each number was obtained  by averaging over $20$ randomly generated network structures.

\begin{table}[ht]
\centering
\footnotesize
\setlength{\tabcolsep}{2.6pt}
\begin{tabular}{@{}ccccccccccccc@{}}
\toprule
	 & \multicolumn{4}{c|}{BA-1}                                                             & \multicolumn{4}{c|}{BA-2}                                                             & \multicolumn{4}{c}{BA-3}                                                             \\ \midrule
\multicolumn{1}{c}{} & \multicolumn{1}{c}{Baseline} & \multicolumn{1}{c}{LESS} & \multicolumn{1}{c}{MINT} & \multicolumn{1}{c|}{MINT-LP} & \multicolumn{1}{c}{Baseline} & \multicolumn{1}{c}{LESS} & \multicolumn{1}{c}{MINT} & \multicolumn{1}{c|}{MINT-LP} & \multicolumn{1}{c}{Baseline} & \multicolumn{1}{c}{LESS} & \multicolumn{1}{c}{MINT} & \multicolumn{1}{c}{MINT-LP} \\ \midrule
1-2-7     & 16.12             & 33.87                 & 13.81                          & \textbf{10.35}             & 21.18                 & 33.54                          & 11.12             & \textbf{10.35}                 & 25.34      & 38.47      & 11.64      & \textbf{10.35}                \\
2-7-1    & 12.08    & 68.46     & \textbf{4.72}     & 7.18       & 15.79        & 84.10       & \textbf{6.18}          & 9.25    & 18.96          & 114.22        & \textbf{7.50}             & 11.77 \\
7-2-1     & 7.12            & 37.18        & \textbf{4.33}       & 6.59           & 8.60         & 62.64         & \textbf{5.60}       & 8.28        & 9.85         & 73.96          & \textbf{6.68}        & 10.25\\
equal    & 11.63                 & 44.24              & \textbf{11.32}              & 17.41        & 15.17           & 65.84            & \textbf{14.98}                    & 22.22        & 18.40                & 70.45                & \textbf{18.35}            & 27.73               
\\ \midrule
& \multicolumn{4}{c|}{SW-1}                                                             & \multicolumn{4}{c|}{SW-2}                                                             & \multicolumn{4}{c}{SW-3}                                                             \\ \midrule
\multicolumn{1}{c}{} & \multicolumn{1}{c}{Baseline} & \multicolumn{1}{c}{LESS} & \multicolumn{1}{c}{MINT} & \multicolumn{1}{c|}{MINT-LP} & \multicolumn{1}{c}{Baseline} & \multicolumn{1}{c}{LESS} & \multicolumn{1}{c}{MINT} & \multicolumn{1}{c|}{MINT-LP} & \multicolumn{1}{c}{Baseline} & \multicolumn{1}{c}{LESS} & \multicolumn{1}{c}{MINT} & \multicolumn{1}{c}{MINT-LP} \\ \midrule
1-2-7   & 29.84                 & 58.55                    & 13.33             & \textbf{10.35}           & 41.97                   & 72.43                  & 10.96            & \textbf{10.35}                &  59.28                   & 23.25                 & 10.96             & \textbf{10.35}        \\
2-7-1     & 25.88                & 40.34             & \textbf{9.68}          & 12.78              &  34.59             & 51.48               & \textbf{13.34}               & 17.34            & 48.67            & 60.97            & 21.02                  & \textbf{19.76}     \\
7-2-1    & 12.10        & 42.08               & \textbf{8.54}      & 11.09    & 15.61              & 51.52               & \textbf{11.66}           & 15.18     & 21.05            & 83.21          & \textbf{16.31}     & 21.74              \\
equal  & \textbf{22.64}            & 56.50            & 23.19                & 29.34                 & \textbf{30.91}           & 53.68          & 36.62            & 33.13     & 43.44             & 52.36           & 36.53             & \textbf{34.50}        \\
\bottomrule
\end{tabular}
\caption{Experiments where \ConfigProbEst and \ConfigProb  were simulated from Spam data.
}\label{tab:spam}
\end{table}

\begin{table}[ht]
\centering
\footnotesize
\setlength{\tabcolsep}{2.6pt}
\begin{tabular}{@{}ccccccccccccc@{}}
\toprule
	 & \multicolumn{4}{c|}{BA-1}                                                             & \multicolumn{4}{c|}{BA-2}                                                             & \multicolumn{4}{c}{BA-3}                                                             \\ \midrule
\multicolumn{1}{c}{} & \multicolumn{1}{c}{Baseline} & \multicolumn{1}{c}{LESS} & \multicolumn{1}{c}{MINT} & \multicolumn{1}{c|}{MINT-LP} & \multicolumn{1}{c}{Baseline} & \multicolumn{1}{c}{LESS} & \multicolumn{1}{c}{MINT} & \multicolumn{1}{c|}{MINT-LP} & \multicolumn{1}{c}{Baseline} & \multicolumn{1}{c}{LESS} & \multicolumn{1}{c}{MINT} & \multicolumn{1}{c}{MINT-LP} \\ \midrule
1-2-7    & 18.75        & 23.28              & 15.06             & \textbf{11.91}                &  23.65             & 33.96               & 12.87              & \textbf{11.90}              &  28.99             & 46.30             & 12.25         & \textbf{11.92}           \\
2-7-1  & 30.04         & 57.11           & \textbf{3.91}           & 5.80           & 36.04            & 79.99              & \textbf{5.04}            & 7.11           &  44.75            & 108.39                    & \textbf{6.20}             & 9.02 \\
7-2-1     & 14.38                  & 56.37                 & \textbf{3.91}         & 5.79              &  16.44               & 48.74              & \textbf{5.05}               & 7.10       & 19.23             & 55.61              & \textbf{6.20}               & 8.97    \\
equal  & 20.25          & 43.70             & \textbf{12.08}     & 18.24            & 24.47           & 51.32              & \textbf{15.39}          & 22.42                & 29.83            & 66.06             & \textbf{18.96}              & 28.04                \\ 
\midrule
& \multicolumn{4}{c|}{SW-1}                                                             & \multicolumn{4}{c|}{SW-2}                                                             & \multicolumn{4}{c}{SW-3}                                                             \\ \midrule
\multicolumn{1}{c}{} & \multicolumn{1}{c}{Baseline} & \multicolumn{1}{c}{LESS} & \multicolumn{1}{c}{MINT} & \multicolumn{1}{c|}{MINT-LP} & \multicolumn{1}{c}{Baseline} & \multicolumn{1}{c}{LESS} & \multicolumn{1}{c}{MINT} & \multicolumn{1}{c|}{MINT-LP} & \multicolumn{1}{c}{Baseline} & \multicolumn{1}{c}{LESS} & \multicolumn{1}{c}{MINT} & \multicolumn{1}{c}{MINT-LP} \\ \midrule
1-2-7    & 31.40           & 39.60             & 12.95           & \textbf{11.92}               & 43.82           & 53.88               & 12.35          & \textbf{11.92}              & 61.62         & 73.53            & 12.06           & \textbf{11.91}    \\
2-7-1      & 51.20           & 41.30          & \textbf{8.22}     & 7.01        & 71.45                & 59.29                 & \textbf{9.77}       & 11.51             & 99.37                   & 90.26             & \textbf{14.02}              & 16.51 \\
7-2-1   & 21.14               & 35.57                & \textbf{6.92}     & 8.17                      & 27.63               & 45.54                  & \textbf{9.59}            & 11.40     & 36.65             & 61.18            & \textbf{13.67}         & 16.31        \\
equal  & 33.15          & 38.67           & \textbf{20.94}          & 25.31           &  45.61               & 52.45             & \textbf{29.04}          & 34.47       & 63.09                & 73.74                 & 43.36           & \textbf{39.72}    \\
\bottomrule
\end{tabular}
\caption{Experiments where \ConfigProbEst and \ConfigProb  were simulated from Hate Speech data.}\label{tab:hate}
\end{table}

The results in  Tables \ref{tab:spam} and \ref{tab:hate} show that the SDP-based (denoted with \ref{eq:MINT}) and LP-based (denoted with \MINTLP) algorithms achieve the best performance except for a few settings. 
In particular, when the penalty of cutting connections between benign nodes is upweighted (i.e., ``2-7-1''), \ref{eq:MINT} consistently outperforms other baseline approaches by a large margin; this shows the importance of considering connectivity when deciding which subset to remove.
In the case of ``1-2-7'', most instances are simple in the sense that they are solvable in polynomial time; as a result, \MINTLP achieves the best performance in these cases; however, \ref{eq:MINT} is not much worse than \ref{eq:MINT-LP}, indicating that the theoretical guarantee in Theorem~\ref{th:guarantee} are  meaningful.
It is worth noting that most instances of ``2-7-1'' are hard in the sense that \ref{eq:MINT-LP} does not directly output integral solutions, instead the perturbation based algorithm was applied. Although the rounding step of the perturbation based algorithm (see step \ref{MINT-LP:op2} in Algorithm~\ref{algo:MINT-LP}) is a heuristic without theoretical guarantee, the losses are not much larger than that of \ref{eq:MINT} in practice.
Finally, the running time of \ref{eq:MINT} and \ref{eq:MINT-LP} is showed in Table~\ref{tab:time-synthetic}.
The experiments were run on a machine with $3.8$GHz CPU.
All results were averaged over the three variants of each network.
For BA networks, \ref{eq:MINT-LP} is on average $42$ times (resp. $5 $ times) faster than \ref{eq:MINT} on Spam (resp. Hate Speech) data; similar speed-up is also observed on Small-World networks.

\begin{table}[ht]
\centering
\begin{tabular}{@{}ccccc@{}}
\toprule
            & \multicolumn{2}{c}{BA} & \multicolumn{2}{c}{Small-World} \\ \midrule
            & \ref{eq:MINT}       & \ref{eq:MINT-LP}   & \ref{eq:MINT}       & \ref{eq:MINT-LP}       \\ \midrule
Spam        & $15.82$s   & $0.37$s   & $14.86$s        & $0.37$s       \\
Hate Speech & $19.00$s   & $3.69$s   & $18.71$s        & $3.66$s       \\ \bottomrule
\end{tabular}
\caption{The average running time of \ref{eq:MINT} and \ref{eq:MINT-LP}. }
\label{tab:time-synthetic}
\end{table}

\subsection{Results for the LP-based Algorithm}
We now present experimental results to demonstrate the scalability of \ref{eq:MINT-LP} on real-world networks. 
We experimented with three real-world networks: 1) a network extracted from Facebook data with $4039$ nodes and $88234$ edges~\cite{leskovec2012learning}; 2) the largest connected component from an email network including $986$ nodes and $16064$ edges~\cite{leskovec2007graph}; 3) the largest connected component from Wikipedia data consisting of $7066$ nodes and $100736$ edges~\cite{leskovec2010predicting}.
For the email network, the distributions \ConfigProbEst and \ConfigProb were simulated from Spam data, while for the other networks the distributions were simulated from Hate Speech data.
The experimental setup is the same as described in the previous section.

\begin{figure}[ht]
\def\FigSize{2.2in}
\centering
\setlength{\tabcolsep}{0.01pt}
\begin{tabular}{ccc}
\includegraphics[width=\FigSize]{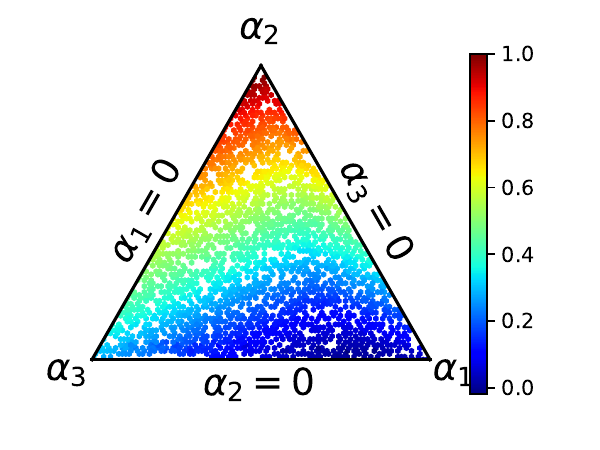} &
\includegraphics[width=\FigSize]{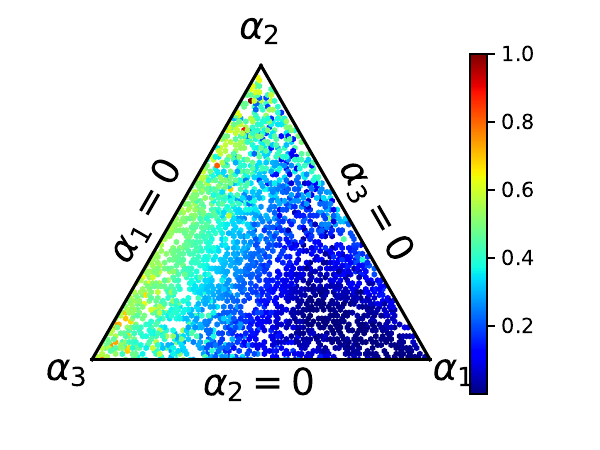} & 
\includegraphics[width=\FigSize]{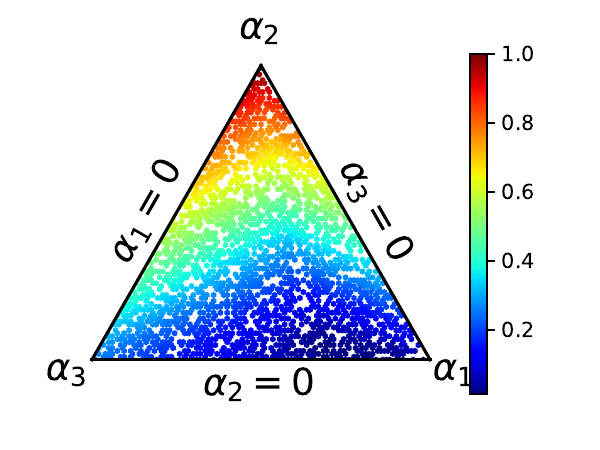}
\end{tabular}
\caption{From left to right: Facebook, Email, and Wikipedia networks. 
Each point represents  $\mathbb{E}_{\ConfigProb}[\Loss_{\text{baseline}}] - \mathbb{E}_{\ConfigProb}[\Loss_{\text{MINT-LP}}]$ normalized across all $3000$ instances. 
}
\label{fig:real-world}
\end{figure}

To have a complete view of how \ref{eq:MINT-LP} performs, for each network, we extensively solved $3000$ instances of $\ref{eq:MINT-LP}(\TradeOffVec, \ConfigProbEst)$ with \TradeOffVec sampled uniformly at random from the standard probability simplex. 
For each instance, we compared with the simple baseline that decides whether or not to remove a node with a threshold , as described in the previous section.
Notice that neither the SDP-based algorithm nor LESS can scale to the real-world networks, due to the enormous memory consumption and high computational costs.
The experimental results are showed in Figure~\ref{fig:real-world}.
Each point in the figure represents the difference of the losses between the simple baseline and \ref{eq:MINT-LP} (i.e., $\mathbb{E}_{\ConfigProb}[\Loss_{\text{baseline}}] - \mathbb{E}_{\ConfigProb}[\Loss_{\text{MINT-LP}}]$), normalized across all $3000$ instances.
It is worth noting that there are only $6$ (out of $3000 \times 3$) instances  on the Facebook network where the simple baseline performs better than \ref{eq:MINT-LP}.
The average running time (over the sampled instances) is:  Email: $10.19$s, Facebook: $122.43$s, Wikipedia: $147.91$s.

\section{Discussion}
    
We considered the problem of removing a subset of nodes from a network such that a customized loss function is minimized. 
In particular, we designed a model named \ref{eq:MINT} that considers both the likelihood that a node is malicious, as well as the network structure. 
Our key insight is for the loss function to capture both the \textit{direct} loss associated with false positives and the \textit{indirect} loss associated with cutting connections between benign nodes, and failing to cut connections from malicious nodes to their benign network neighbors. 
We first showed that optimally solving \ref{eq:MINT} is NP-hard, even under very restrictive settings, e.g., the nodes' probabilities of being malicious are independent.
Due to the hardness results, we proposed an approximation algorithm based on Semidefinite programming (SDP) and randomized rounding.
We also provided theoretical bound to guarantee the quality of approximation.
Experimental results showed that the approximation algorithm outperforms alternative approaches in terms of loss.

Although the SDP-based approximation algorithm enjoys both nice approximation guarantees and good empirical performance, it may be limited by the scalability to large-scale networks.
The approximation algorithm consists of two steps: 1) solving an SDP for an optimal solution in continuous space and 2) converting the optimal solution back to discrete space with a randomized rounding algorithm.
The bottle neck is solving the SDP, which costs $O(\NumAgent^{3.5})$.
To improve the scalability, we reformulated \ref{eq:MINT} as a linear programming (LP), i.e, \ref{eq:MINT-LP}; empirically, the LP is much cheaper to solve than solving the SDP.
Intuitively, instead of solving the SDP in step 1, we solved the \ref{eq:MINT-LP}.
One interesting observation is that the LP formulations of many instances admitted integral solutions despite the NP-hardness; indeed, worst-case instances may be rare in practice; as a result, \ref{eq:MINT-LP} directly outputs optimal solutions of \ref{eq:MINT}.
For those really hard instances (i.e., the LP formulations do not admit integral solutions), we proposed a simple heuristic algorithm to convert the fractional solutions back to discrete domain.

Currently, we assume that the maliciousness probability distribution \ConfigProbEst is given as an input to \ref{eq:MINT}; the distribution can be estimated by a suitable ML model, e.g., graph neural networks.
However, the training of the ML model usually focuses on minimizing prediction accuracy, which is at most a proxy of the actual loss that we care about.
One future direction is to unify the training of the ML model and the optimization of \ref{eq:MINT} in an end-to-end fashion, such that the training is directly supervised by the decision quality.


\newpage
\bibliography{main}
\bibliographystyle{abbrvnat}

\end{document}